\documentclass[amsmath,amssymb,preprint]{revtex4}

\usepackage{graphicx}
\usepackage{dcolumn}
\usepackage{bm}
\begin{document}

\title{Stability of the Bloch wall via the Bogomolnyi decomposition
in elliptic coordinates}

\author{S.R. Woodford}
\altaffiliation{On leave of absence from Forschungszentrum J\"ulich.
Postal address:
Theorie I, Institut f\"ur Festk\"orperforschung,
Forschungszentrum J\"ulich,  D-52428 J\"ulich, Germany}
 \email{s.woodford@fz-juelich.de}

\author{I.V. Barashenkov}
 \email{Igor.Barashenkov@uct.ac.za; igor@odette.mth.uct.ac.za}
\affiliation{Department of Mathematics and Applied Mathematics,
University of Cape
Town, Rondebosch 7701, South Africa}

\date{\today}

\begin{abstract}
 
We consider the one-dimensional
anisotropic $XY$ model in the continuum limit.
  Stability analysis of its Bloch wall solution 
  is hindered by the nondiagonality of the 
  associated linearised
  operator and the hessian of energy. 
  We circumvent this 
difficulty  by showing that the energy 
admits a Bogomolnyi bound in elliptic
coordinates and that the Bloch wall saturates it ---
that is, the Bloch wall renders the energy minimum.
Our analysis provides a simple but
nontrivial application of the BPS (Bogomolnyi - Prasad - Sommerfield)
construction in one 
dimension, where its use is often believed to be limited to  reproducing 
results obtainable by other means. 

\end{abstract}


\maketitle

\section{Introduction}

Although the Bogomolnyi
decomposition is an indispensable tool for the study of 
higher-dimensional field theories \cite{BPS}, it has seldom been used in
one dimension. It does  allow one to reduce the order 
and find topological solitons of one-component models such as the 
sine-Gordon and $\phi^4$-theory, but these can be obtained with less
effort simply by using an integrating factor. It is of more value for
multicomponent \cite{Bazeia} 
and lattice \cite{SW} systems, but all models that benefited
from its use 
had to
be specifically designed to admit such a decomposition.

   In this paper, we apply the Bogomolnyi construction to a system 
that has been studied for more than forty years: the anisotropic $XY$\ model. 
It has long been believed that the Bloch-wall solutions of 
this model are stable; this fact has been demonstrated numerically 
 but   never proven analytically. 
Here, by transforming to elliptic coordinates, we 
find the Bogomolnyi bound 
for the energy and show that the Bloch wall  minimizes the  energy in
the corresponding  
topological sector; this  proves its stability. (Note
that the Bogomolnyi construction cannot be carried out in the original
coordinates.) 

An outline of the  paper is as follows. In the next section
we introduce the (stationary) anisotropic
$XY$ model 
and its domain-wall solutions. We also describe
three different time-dependent extensions of the model which correspond
to
three
different types of behaviour, {\it viz.}, dissipative,
conservative relativistic and conservative
nonrelativistic dynamics. In section \ref{stab} we 
show that because of the nondiagonality of the 
linearised operator and the second variation of energy,
the standard methods  are inadequate for proving the stability of
the Bloch wall --- in each of the three cases.
Section \ref{Bogo} contains the main result of this paper: the 
proof 
of the energy minimization
by the Bloch wall.
Finally, section \ref{Conc} summarises the results of our work.

\section{The model}
\label{model}

\subsection{Anisotropic easy-axis ferromagnet near the Curie temperature}

The anisotropic  $XY$\ model
was originally introduced  to describe domain walls in an easy-axis 
ferromagnet near the Curie point 
\cite{Ginsburg,Lajzerowicz2}.  
 In the study of domain walls, 
the magnetization is assumed to vary only in $x$\
direction and the magnetization vector ${\bf M}$ is taken to lie in
the $yz$\ plane.
(With this  notation, 
the system  should have been called the $YZ$\ model
but we keep the traditional name to avoid confusion.)
In the continuum limit, the model is defined by its free energy expansion
\cite{Ginsburg,Lajzerowicz2}: 
\begin{equation}
E = \int \left[
\frac 12 (\partial_x \mathbf{M})^2 - (1+h)\mathbf{M}^2 
+ \frac 12 \mathbf{M}^4 + 2h M_y^2 + \mathcal{E}_0
\right] dx.
\label{Free}
\end{equation}
The first term in \eqref{Free} is the exchange energy, 
which is minimized when ${\bf M}=const$. 
The combination
of the second and third terms makes 
the $|{\bf M}| \neq 0$  ground state energetically preferable 
to the state with $|{\bf M}| = 0$. Anisotropy is caused by
the term $2h M_y^2$; 
the  parameter $h$\ is assumed positive, 
making  the $z$\ direction 
the easy-axis. Finally, the coefficient $(1+h)$\ is 
introduced to simplify the subsequent formulas
and a constant $\mathcal{E}_0$\ has been added  
to the integrand to ensure that the  free energy 
 is finite.

Defining  $\psi = M_y + iM_z$, 
we can recast the free energy  in the form 
\begin{equation}
E = \int \left[  \frac 12 |\psi_x|^2 + \frac 12 |\psi|^4 - |\psi|^2 
+ \frac{h}{2} (\psi^2+ \psi^{*2})  + \mathcal{E}_0 \right]dx.
\label{density}
\end{equation}
This is the form that we shall be working with in this paper.

\subsection{Anisotropic easy-plane ferromagnet in external field} 
\label{easy-plane}

The same free energy expansion (\ref{density})  
describes one more, unrelated,
magnetic system:
 a weakly anisotropic {\it easy-plane\/} ferromagnet in an external  
 field  perpendicular to the easy plane \cite{BWZ}. 
 In terms of the magnetization vector $\mathbf{M}$, 
 the free energy  of this system is given by
\begin{equation}
E = \int \left[ \frac{1}{2} (\partial_x \mathbf{M})^2 + 
\frac{\beta}{2} M_z^2 + \frac{\epsilon \beta}{2}M_x^2 -HM_z
+ \mathcal{E}_0 \right] dx, 
 \label{Free2}
\end{equation}
where $H>0$ is an applied
field, 
 the anisotropy parameters $\beta $ and $\epsilon \beta $
are positive, and $\epsilon \ll 1$.
In this model, the $xy$\ 
plane is the easy plane, with a weak anisotropy favouring the $y$\
 direction.  Far from the Curie point,  the magnetization is hard,
 i.e. the magnitude of the magnetisation vector
 is  constant: $\mathbf{M}^2 = M_0^2 = \mbox{const}$. 
 [For this reason equation (\ref{Free2}) has no terms in $\mathbf{M}^2$\ and 
 $\mathbf{M}^4$.]
 
We assume that the  the magnitude of the external magnetic
field is  close to $\beta M_0$: 
$H = \beta M_0 - \epsilon q$,
where $q$ is a quantity of order 1.  
The (degenerate) ground states will then
have the magnetization vector almost parallel to the field:
\[
\mathbf{M}^{(\pm)} \approx (0,  \pm \sqrt{2\epsilon q M_0/\beta}, M_0 
- \epsilon q/\beta). 
\]
The deviation from the reference vector
$\mathbf{M} = (0,0,M_0)$\ pointing in the
direction of the field can be characterised
using the complex variable $\psi = (\beta / 2\epsilon s)^{1/2} (M_x - i M_y)$, 
where $s \equiv qM_0 - \beta M_0^2 /2$. The $M_z$ component of
the magnetisation is expressible using 
$\mathbf{M}^2 = M_0^2 = \mbox{const}$: 
\[
M_z =M_0 \sqrt{1-\frac{2 \epsilon s}{ \beta M_0^2} |\psi|^2}.
\]
Transforming to ${\tilde x}=  (\epsilon s/2 \alpha M_0^2)^{1/2}x$
and keeping terms up to $\epsilon^2$, 
the free energy  (\ref{Free2}) reduces to equation (\ref{density}) where
$h = \beta M_0^2/2s$  and we have  dropped the tilde
over $x$. Unlike the anisotropic $XY$\ model considered in
the previous subsection, 
the domain walls in the present system interpolate between the 
nearly-parallel ground states $\mathbf{M}^{(+)}$
and $\mathbf{M}^{(-)}$. It is fitting to note here that 
due to the  relative  weakness of  magnetic anisotropy, 
the field $H \approx \beta M_0$\ will generally be well 
within the range of modern experiments. For example, the martensite 
phase of {\rm NiMnGa} is a weakly anisotropic easy-plane ferromagnet
with $\beta \approx 2 \times 10^{-6} \text{Tm/A}$ and $M_0 \approx 5
\times 10^{5} \text{A/m}$ 
\cite{Straka}. Therefore, when subjected to a magnetic
field 
$H \approx   1 \text{T}$, it will be described by 
equation (\ref{density}).

Finally, we note that the expansion
\eqref{Free}-\eqref{density}  
arises in yet another magnetic context, {\it viz.}
a strongly anisotropic ferromagnet with spin $S=1$
\cite{Ivanov}.

\subsection{Bloch and Ising walls} 

The free energy \eqref{density} is extremised by solutions
of the following stationary equation:
\begin{equation}
\frac{1}{2} \psi_{xx} -|\psi|^2 \psi + \psi - h \psi^*=0.
\label{static}
\end{equation}
There are two soliton, or kink,  solutions available in literature;
each describes an interface between two ferromagnetic domains.
One is commonly known as the Ising, or N\'eel, wall 
\cite{Ginsburg,Lajzerowicz2,Raj_Wei,EM}:
\begin{equation}
\psi_I(x)= i A \tanh (Ax),
\label{Ising}
\end{equation}
where $A=\sqrt{1+h}$.
The second kink solution has the form
\begin{equation}
\psi_B(x)= i A \tanh (Bx) \pm C \mbox{sech} (Bx),
\label{Bloch}
\end{equation}
where $B= \sqrt{4h}$ and $C=\sqrt{1-3h}$ 
\cite{Lajzerowicz2,Sar_Tru_Bish,Montonen}.
This solution is referred to as the Bloch wall. 
The two signs in front of the real part in \eqref{Bloch} distinguish 
Bloch walls of opposite chiralities.
The Ising wall exists for all positive $h$ whereas the Bloch wall exists
only for $h < \frac13$.
 
The Bloch and Ising wall have the same asymptotic behaviour: $\psi_{B,I}(x)
\to \pm iA$ as $x \to \pm \infty$. This determines the constant ${\cal E}_0$
that is added to the integrand in \eqref{density} to ensure 
the finiteness of  the free energy: ${\cal E}_0=A^4/2$.
The Bloch and Ising walls have energies 
\begin{equation}
\label{Eb}
E_B = 2B(A^2 - B^2/3)
\end{equation}
 and $E_I = \frac43 A^3$, respectively. 
It is easily verified 
that the energy of the Ising wall is greater than the energy 
of the Bloch wall for all $h< \frac 13$. 

In addition to the Bloch and Ising walls,
equation \eqref{static} has a family of nontopological
solitons, available in explicit form  and
describing Bloch-Ising bound states \cite{BW}.
These will not be considered here. 

\subsection{Three types of dynamical behaviour}

The dynamics of the domain walls is governed by one
of three possible  time-dependent extensions
of equation \eqref{static}.
In the case when the free energy \eqref{density} is
used to model the ferromagnet near the Curie point,
the evolution of the field $\psi$ is
dissipative and governed  by
the 
Ginsburg-Landau equation \cite{GL_mag}:
\begin{equation}
\label{GL}
\psi_t =\psi - |\psi|^2 \psi+ \frac 12 \psi_{xx}    - h\psi^*.
\end{equation}

On the other hand, 
in the case of the ferromagnet with spin $S=1$,
the field $\psi$ satisfies a relativistically-invariant equation \cite{Ivanov}
\begin{equation}
\psi_{tt} -  \psi_{xx} -2\psi  + 2|\psi|^2\psi  + 2h\psi^{*} = 0.
\label{MSTB}
\end{equation}
Originally, this equation was  introduced by
  Montonen \cite{Montonen} in a different context  --- as an
exactly solvable special case of 
Rajaraman and Weinberg's bag model \cite{Raj_Wei}. Independently,
 Sarker, Trullinger and Bishop \cite{Sar_Tru_Bish}
proposed it as an interesting interpolate 
 between the sine-Gordon  and the $\phi^4$
theories. Accordingly, the Klein-Gordon equation \eqref{MSTB} is commonly
known as the Montonen-Sarker-Trullinger-Bishop (MSTB) model.

Finally, the magnetisation vector of an
anisotropic easy-plane ferromagnet in an external
field satisfies the Landau-Lifshitz equation. 
The perturbation procedure  described
in section \ref{easy-plane} reduces it \cite{BWZ} to
the 
parametrically driven nonlinear Schr\"odinger equation,
\begin{equation}
i \psi_t+ \frac12 \psi_{xx} -\psi|\psi|^2 + \psi= h \psi^*.
\label{NLS}
\end{equation}

The stability of the Bloch or Ising wall 
depends  on which of the three equations \eqref{GL}, \eqref{MSTB}
or \eqref{NLS} governs the evolution of $\psi$.
Consequently, the three cases --- the Ginsburg-Landau, the Klein-Gordon and the 
nonlinear Schr\"odinger --- need to be considered 
 separately. 

\section{Approaches to stability}
\label{stab}

In this
section we describe the standard methods of stability analysis
and explain why they all fail in the case of the 
Bloch wall --- for each of the three types of evolution.

\subsection{The Ginsburg-Landau and relativistic dynamics}

We start with the Ginsburg-Landau equation, equation \eqref{GL},
and linearise it  about the stationary solution 
$\psi_0(x)$, which can be either the
Bloch or Ising wall. Decomposing the small perturbation $\delta \psi(x,t)$
into its real and imaginary parts, $\delta \psi(x,t)=\delta {\cal R}(x,t)
+ i \delta {\cal I} (x,t)$, and letting 
\[
\delta {\cal R}(x,t)= \tilde{\delta {\cal R}}(x)e^{\lambda t}, \quad
 \delta {\cal I}(x,t)= \tilde{\delta  {\cal I}}(x)e^{\lambda t},
 \]
yields an eigenvalue problem 
\begin{equation}
- {\cal H} \left(
\begin{array}{c} \delta {\cal R} \\ \delta {\cal I}
\end{array}
\right)= \lambda \left(        
\begin{array}{c} \delta {\cal R}  \\ \delta {\cal I} 
\end{array}
\right),
\label{EV}
\end{equation}
where $\cal H$ is a self-adjoint operator
\begin{equation}
{\cal H}=
\left(
\begin{array}{lr}
-\frac12 \partial_x^2 -1+h + 3{\cal R}_0^2 + {\cal I}_0^2 &
2 {\cal R}_0 {\cal I}_0 \\ 2 {\cal R}_0 {\cal I}_0
& -\frac12 \partial_x^2 -1-h + {\cal R}_0^2 + 3{\cal I}_0^2
 \end{array}
\right).
\label{H}
\end{equation} 
In \eqref{H}, ${\cal R}_0$ and ${\cal I}_0$ stand for the real 
and imaginary part of the stationary solution $\psi_0$, 
respectively.
The solution $\psi_0$  will be unstable if the operator
$- {\cal H}$ has at least one 
positive eigenvalue $\lambda$, and stable otherwise.
  
In the case of the MSTB model [equation \eqref{MSTB}],
the linearisation about $\psi_0$ produces the same eigenvalue problem 
\eqref{EV}, with the same operator \eqref{H},
 where one just needs to replace $\lambda$ with $\lambda^2/2$.
Here we have the same stability criterion as in the 
Ginsburg-Landau case: the solution $\psi_0$  will be unstable
if  $-{\cal H}$ has at least one positive eigenvalue
$\lambda^2/2$.

An alternative approach 
(which still leads to the same criterion, though) 
is based on considering the Lyapunov
functional (see e.g. \cite{Lyapunov1}). The Ginsburg-Landau
equation \eqref{GL} can be written as $
\psi_t = -\delta E / \delta \psi^*$,
where $E$ is the functional \eqref{density}.
Hence this functional  satisfies
\[
 E_t= -2 \int |\psi_t|^2 dx,
\]
and so we have $E_t <0$ unless $\psi$ is a static solution, $\psi_t=0$. 
Now if we could prove that $E[\psi]>E[\psi_0]$ for all $\psi$ in
some neighbourhood of $\psi_0$,
the functional $E$ would be the Lyapunov functional 
for this solution and hence
 $\psi_0$ would be proven to be stable.

Here we need to make a standard remark on the translational invariance
of equations \eqref{GL}, \eqref{MSTB} and \eqref{NLS}. 
The domain wall centred at the
origin has the same energy as the wall centred at any other point $x_0$
and therefore can never be an isolated minimum of $E$. However, a 
mere translation of the wall from the origin to the point $x_0$ does not
imply instability. Therefore, it is physically reasonable
to group all configurations
obtained from a given $\psi(x)$ by translations $\psi(x) \to
\psi(x-x_0)$ with $-\infty< x_0 < \infty$, into 
equivalence classes. Unlike the wall with any particular $x_0$,
 the equivalence class consisting of all translated walls
{\it can\/} be an isolated minimum of the energy 
--- defined on the corresponding
quotient manifold.
If the perturbations of the wall are assumed to be infinitesimal,
the quotient is a linear subspace; namely,
  it is the quotient of the space of 
all infinitesimal perturbations of the wall
by the subspace spanned by its translation mode
$(\partial_x{\cal R}_0, 
\partial_x{\cal I}_0)$.
This quotient space can be conveniently characterised 
by the orthogonality constraint 
\begin{equation}
\int (\partial_x{\cal R}_0, 
\partial_x{\cal I}_0 ) 
\left( \begin{array}{c} \delta {\cal R} \\ \delta {\cal I} 
\end{array} \right) dx=0.
\label{constraint}
\end{equation} 
In what follows, the subspace of perturbations defined by the constraint
\eqref{constraint} will be denoted ${\cal S}$.

To check whether $\psi_0$ renders $E[\psi]$ a minimum
in $\cal S$, 
the functional is expanded about the stationary point $\psi_0$:
\begin{equation*}
E[\psi]= E_0 + \frac12 \delta^2 E + ... .
\label{expansion}
\end{equation*}
Here $\psi=\psi_0 + \delta {\cal R} + i \delta {\cal I}$, and
the second variation has the form
\begin{equation}
\frac12 \delta^2 E= \int (\delta {\cal R}, \delta {\cal I}) \, 
{\cal H} \left(
\begin{array}{c} \delta {\cal R} \\ \delta  {\cal I}
\end{array}
\right) dx,
\label{star}
\end{equation}
where the hessian $\cal H$  coincides with  the linearised operator \eqref{H}.
The solution $\psi_0$ will
minimise $E[\psi]$ in $\cal S$ provided $\cal H$ has no negative 
or zero eigenvalues other than the one associated with the
translation mode.

A Lyapunov functional can also be used in the case of the
relativistic dynamics \eqref{MSTB} (see e.g. \cite{Lyapunov2}).
Here, as a candidate functional one considers the total
energy
\begin{equation}
E_{\rm total} \, [\psi, \psi_t]=  \frac12 \int |\psi_t|^2 dx+ E[\psi],
\label{E_rel}
\end{equation}
where  $E[\psi]$ is as in \eqref{density}. The energy is 
conserved, and hence
 $ E_{\rm total}$ will define the Lyapunov functional 
for the solution $\psi_0$ if it renders this functional a minimum
in $\cal S$.
Since the first term in \eqref{E_rel} is minimised
by any static configuration, it is sufficient to check
whether $\psi_0(x)$ renders the functional $E[\psi]$ a minimum.
Consequently, 
the minimisation problem reduces to the
 eigenvalue problem \eqref{EV}.

Thus, we have the same stability criterion in the 
case of the Ginsburg-Landau and relativistic dynamics ---
one has to  prove that the solution minimises the 
functional \eqref{density} under the constraint \eqref{constraint},
or, equivalently, show that 
the operator $ {\cal H}$ has no negative 
or zero eigenvalues other than the translation mode.

In the case of the Ising wall \eqref{Ising}, we have ${\cal R}_0(x)=0$ 
and the operator \eqref{H} is
diagonal. Its eigenvalues can be readily found and the above 
  criterion easily 
implemented. Thus, it was shown in 
\cite{Raj_Wei,Lajzerowicz1,Ito_Tasaki,Ivanov} 
 that the Ising wall 
is stable for $h>\frac13$ and unstable for $h<\frac13$. 
(See \cite{Skryabin} for the generalisation to a 
nonvariational case.)

In the case of the Bloch wall, 
on the other hand, the operator $\cal H$ is 
nondiagonal. This makes it impossible to determine the sign of the lowest eigenvalue
 of $\cal H$ using
standard analytical methods. Consequently, 
previous studies had  to resort to 
semi-intuitive and numerical arguments. In particular,
it was noted  
that the energy of the Bloch wall is lower than that of 
the Ising wall and suggested 
that the Bloch wall should be stable \cite{Montonen,Sar_Tru_Bish,Sub_Tru}. 
This conjecture was supported
by results of  direct numerical simulations of equation \eqref{MSTB}
\cite{Haw_Sab_Tru}
and perturbation theory for small $C$ ($C=\sqrt{1-3h}$)
\cite{Ivanov}. However no analytical proof, applicable for 
all  parameter values, has  been given so far.

\subsection{The Schr\"odinger dynamics}

Finally, we discuss the 
parametrically driven nonlinear Schr\"odinger equation,
equation \eqref{NLS}.
In this case, the linearisation about $\psi_0$ produces
a symplectic eigenvalue problem
\begin{equation}
{\cal H} \left(
\begin{array}{c} \delta {\cal R} \\ \delta {\cal I}
\end{array}
\right)= \lambda J 
 \left(
\begin{array}{c} \delta {\cal R}  \\ \delta {\cal I} 
\end{array}
\right), \quad J= \left(  \begin{array}{rr}
0 & -1 \\ 1 & 0             
\end{array} \right),
\label{EV_NLS}
\end{equation}
where $\cal H$ is as in \eqref{H}. 
The product operator $J^{-1} {\cal H}$ is 
non-self-adjoint and hence its eigenvalues 
can be complex.
The solution $\psi_0$ will be unstable if the operator
$J^{-1} {\cal H}$ has at least one eigenvalue 
 $\lambda$ with  positive real part. 

The nonlinear Schr\"odinger 
equation \eqref{NLS} conserves energy which is given by the integral
\eqref{density}; hence the energy is  a potential Lyapunov 
functional for
equation \eqref{NLS}.
One can therefore try to establish the stability of $\psi_0$
by proving that it renders the energy minimum
under the constraint \eqref{constraint}; this 
happens when the operator $\cal H$ in the second variation \eqref{star}
does not have negative or zero eigenvalues other than the translation
mode.
Note that the criteria based on the linearisation and 
the energy minimality appeal to eigenvalues of {\it different\/} operators
here, $J^{-1} \cal H$ and
$\cal H$, respectively. However, 
it is not difficult to show that the positive definiteness
of $\cal H$ {\it implies\/} that $J^{-1} \cal H$ does not have
eigenvalues with  positive real part. We include a proof
of this simple fact in the Appendix.

As we have  pointed out in the previous
subsection,  the operator $\cal H$
associated with the Ising wall is diagonal.
 Making use of  this property, it was proved in \cite{BWZ} that
 $J^{-1} {\cal H}$ does not have  eigenvalues
 with positive real part and
the Ising wall is stable for all $h>0$. On the other hand, in the case
of the Bloch wall, 
 the operator $\cal H$ in \eqref{EV} and \eqref{EV_NLS}
is nondiagonal. This prevents the determination of
the sign of
the lowest eigenvalue of $\cal H$, or testing
the existence of unstable eigenvalues of $J^{-1} \cal H$,
using any of the
standard analytical approaches.
The 
eigenvalue problems \eqref{EV} and \eqref{EV_NLS} can 
of course be 
studied numerically; this was done in 
Ref.\cite{BWZ} where the Bloch wall was found to be stable 
for all examined values of $h$.
However,
numerical solutions tend to overlook subtleties (e.g. exponentially small
eigenvalues) and give limited insights into the 
structure of the configuration space.
This motivates our search for an analytical stability proof.

We provide such a proof in the next section.

\section{$XY$ model in elliptic  coordinates}
\label{Bogo}

Returning to  the energy (\ref{density}), we
define elliptic coordinates on the ($\text{Re} \, \psi$,
$\text{Im}\, 
\psi$)-plane:
\begin{equation}
\psi = B \left(\sinh{u} \sin{v} + i \cosh{u}\cos{v} \right).
\label{trans}
\end{equation}
Here, $u(x) \geq 0$\ and $0 \leq v(x) \leq 2\pi$\ are continuous fields.

The use of elliptic coordinates 
 was pioneered by Trullinger and  DeLeonardis
who utilised these
in their calculation of the partition function 
for the MSTB model
\cite{Tru_DL}. Elliptic coordinates allow the separation
of variables in
the effective Schr\"odinger equation that arises in their transfer-matrix 
approach 
(see also \cite{Sub_Tru}). Subsequently, Ito \cite{Ito} used
 elliptic coordinates
to separate   variables in the Hamilton-Jacobi formulation of equation
\eqref{static} 
(see also \cite{Guilarte_98}).

 Transforming to the elliptic coordinates (\ref{trans}), the energy
 functional \eqref{density} acquires the form
\begin{equation}
E = 2h \int 
\left[ (\sinh^2{u} + \sin^2{v}) (u_x^2 + v_x^2) + 
\frac{f^2(u) + g^2(v)}{\sinh^2{u} 
+ \sin^2{v}} \right] dx,
\label{E2h}
\end{equation}
where 
\begin{subequations}
\label{fg}
\begin{eqnarray} 
f(u) = B  
\sinh{u} \left(\cosh^2{u} - \frac{A^2}{B^2} \right), \label{f} \\
g(v) = B 
 \sin{v} \left(\frac{A^2}{B^2} -\cos^2{v} \right). \label{g}
\end{eqnarray}
\end{subequations} 
 The integrand in \eqref{E2h} admits a Bogomolnyi-type decomposition
\begin{equation}
E = 2h \int \left\{\mu(u,v)
\left[\left(u_x + \frac{f(u)}{\mu(u,v)}\right)^2 +
 \left(v_x + \frac{g(v)}{\mu(u,v)}\right)^2 \right] + \Phi_x \right\}
 dx, 
\label{BPS2}
\end{equation} 
where 
\[
\mu(u,v)= \sinh^2 u + \sin^2 v \geq 0
\]
and
 \begin{equation*}
 \Phi_x = -2f(u)u_x - 2g(v)v_x.
\label{Phi}
\end{equation*}
Since   both terms in the square brackets in 
\eqref{BPS2} are nonnegative, the energy is bounded from below: 
\begin{equation}
\left. \phantom{\frac12} E \geq 2h \Phi(x) \right|^{\infty}_{-\infty}.
\label{ineq}
\end{equation}
Evaluating the right-hand side of \eqref{ineq} using
\eqref{fg}, this inequality is transformed 
into
\begin{eqnarray}
E \geq  \left. \frac{B}{3} 
\left\{ \left[3A^2 \cosh  u(x) -
B^2\cosh^3 u(x)   
\right] 
+ \left[3A^2 \cos v(x)-
B^2 \cos^3 v(x)  \right]  \right\} \right|_{-\infty}^{\infty}.
\label{Bound}
\end{eqnarray} 

In terms of the 
elliptic coordinates, the domain walls' boundary conditions
  $\psi(\pm\infty) = \pm iA$ acquire the form
  \begin{subequations} \label{BC}
  \begin{eqnarray}
 u(-\infty) = \mbox{arccosh} \left( \frac{A}{B} \right), 
 \quad 
 v(-\infty) = \pi; 
 \label{left} \\
   u(+\infty) = \mbox{arccosh}\left( \frac{A}{B}\right),
   \quad v(+\infty)=0,
   \label{right1} \end{eqnarray}
   or
   \begin{equation}
   u(+\infty) = \mbox{arccosh}\left( \frac{A}{B}\right),
   \quad v(+\infty)= 2 \pi.
   \label{right2}
   \end{equation}
 \end{subequations}
 Using these, the inequality \eqref{Bound} 
becomes simply 
\begin{equation}
E[\psi] \geq E_B,
\label{Bound2}
\end{equation}
where $E_B$ is the energy of the Bloch 
wall given by \eqref{Eb}.

 The bound (\ref{Bound2}) is obviously saturated by the Bloch walls
 \eqref{Bloch}. 
 We now show that, given the boundary 
conditions \eqref{BC},
  the two Bloch walls are the {\it only\/}
solutions with the minimum energy.  
The proof appeals to  
the Bogomolnyi equations
\begin{subequations} 
\label{Bogomolny} 
\begin{eqnarray}
u_x = -\frac{B}{\mu(u,v)}   
\sinh{u} \left(\cosh^2{u} - \frac{A^2}{B^2} \right),\label{Eq1}
 \\
v_x = -\frac{B}{\mu(u,v)} 
 \sin{v} \left(\frac{A^2}{B^2} -\cos^2{v} \right), \label{Eq2}
\end{eqnarray} 
\end{subequations}
which have to be satisfied by any configuration with
 $E[\psi]=E_B$.
 In order to prove the uniqueness, it is sufficient
to demonstrate that the dynamical system \eqref{Bogomolny} 
has a unique heteroclinic trajectory connecting the 
point \eqref{left} to the point \eqref{right1}
and another unique trajectory connecting \eqref{left} to  \eqref{right2}.
To this end, we note that 
 the line $u =  \mbox{arccosh}(A/B)$
is an invariant manifold and that this manifold is
 attractive: trajectories flow towards this line but
 no trajectories can leave it.
Therefore, the only trajectories connecting the  points
\eqref{BC} have to be segments of this straight line. 
 Letting $u =  \mbox{arccosh}(A/B)$ , equation \eqref{Eq2} simplifies to 
\begin{equation*}
v_x = -B \sin v.
\label{simple}
\end{equation*}
Subject to the boundary conditions 
$v(-\infty) = \pi$, $v(\infty) = 0$ and 
$v(-\infty) = \pi$, $v(\infty) =2\pi$,  this equation  
has a unique  pair of solutions 
\begin{equation}
\sin v = \pm \mbox{sech}(Bx), \quad \cos v = \mbox{tanh}(Bx).
\label{Solution}
\end{equation}
(More precisely, these solutions are unique up 
to translations $x \to x-x_0$.) Inserting 
equations \eqref{Solution} into  (\ref{trans}) yields the right- and 
left-handed Bloch walls, equation (\ref{Bloch}). 
Consequently, the Bloch walls are indeed 
the unique minimal energy solutions (modulo translations). 
This proves their stability --- within each of the three evolution
equations \eqref{GL}, \eqref{MSTB} and \eqref{NLS}.

\section{Concluding remarks}
\label{Conc}

We have utilised the Bogomolnyi construction to prove that
the Bloch walls of equation \eqref{density}
 are energy-minimizing kinks. We have also shown
 that they are {\it unique\/} energy minimizers. Thus for all three
evolution equations \eqref{GL}, \eqref{MSTB}, and
\eqref{NLS}, the Bloch walls are proven to be stable.
                                                                                
The energy \eqref{density} was introduced 
 to describe the anisotropic $XY$ model.
However, the related evolution equations 
\eqref{GL}, \eqref{MSTB} and \eqref{NLS} emerge in
several other areas where our results will also be applicable.
In particular, the Ginsburg-Landau equation \eqref{GL}
appears as a generic amplitude equation in resonantly
forced oscillatory
media  near the Hopf bifurcation \cite{GL_gen}.
The Bloch and Ising walls are often regarded as 
the basic building blocks
for the  one-
and two-dimensional patterns arising in such media \cite{BI_gen}. 
Our stability result puts this interpretation on a
firmer ground. 
Next, the MSTB model \eqref{MSTB} was studied in the context of quantum field 
theory \cite{Guilarte_88}. 
Here, the fact that the model admits a BPS
 bound is
of fundamental importance as it means that it admits a natural
supersymmetric extension.
Finally, 
the parametrically driven NLS equation \eqref{NLS}
describes Faraday resonance in a wide shallow water channel 
in the low-viscosity limit
 \cite{EM,NLS_fluid}. 
Since the Ising wall has already been 
shown to be stable within the
NLS equation \cite{BWZ}, a similar
conclusion obtained now for the Bloch wall
reveals an interesting bistability of the two solitons. 
This bistability should allow experimental realization.

\acknowledgments

It is a pleasure to thank Alexander Ianovsky and Dmitry Pelinovsky
for useful remarks.
S.W. was supported by a grant from the Science Faculty of 
the University of Cape Town.
I.B. was supported by the NRF of South Africa under
grant 2053723.

\appendix
\section{The stability-minimality 
correspondence for the nonlinear Schr\"odinger dynamics}

While the linearised operator coincides with the
hessian of energy in the case of the Ginsburg-Landau
and relativistic dynamics,
the two operators are different 
 in the nonlinear Schr\"odinger case. 
 The hessian $\cal H$ is given by equation \eqref{H},
 whereas 
 the linearised operator 
is  $J^{-1} {\cal H}$, with  $J$ 
the skew-symmetric 
matrix \eqref{EV_NLS}. The multiplication by a skew-symmetric matrix 
changes the spectral properties of an operator; 
for example, the continuous spectrum of $\cal H$ lies on the 
positive real axis whereas the continuous spectrum of
$J^{-1} {\cal H}$ consists of pure imaginary $\lambda$.
Therefore it is not 
obvious how the energy minimality translates into the absence
of unstable eigenvalues.

In this appendix we provide a simple proof  that
the absence of negative 
and zero eigenvalues of 
 $\left. \phantom{\frac12} \cal H \right|_{\cal S}$ is sufficient
for $J^{-1} {\cal H}$ not to have eigenvalues with 
positive real part. This fact is usually familiar
to workers in this field;
for the comprehensive treatment, including the
necessary conditions for stability and the eigenvalue count, see
\cite{E_vs_stability}.

First of all, we note that if $\lambda$ is an eigenvalue 
of the operator $J^{-1} {\cal H}$ 
 in \eqref{EV_NLS}, associated with an eigenvector
\[ {\bf z}(x)= \left( \begin{array}{c} a(x) \\ b(x) \end{array} \right),
\]
then $-\lambda$ is also an eigenvalue, associated
with the eigenvector 
\begin{equation}
{\tilde {\bf z}}(x)= \left( \begin{array}{c} a(-x) \\ -b(-x) \end{array} 
\right).\label{tilde}
\end{equation}
[This follows from the fact that both for the Bloch and Ising wall,
the off-diagonal elements of the matrix $\cal H$ in \eqref{H}
are odd functions of
$x$:  ${\cal H}_{12}(x)={\cal H}_{21}(x)=-{\cal H}_{21}(-x)$.]
Hence real eigenvalues always come in $(\lambda, -\lambda)$ pairs. 
On the other hand, if $\lambda$ is a complex eigenvalue with
an eigenvector ${\bf z}(x)$, then $\lambda^*$ is an eigenvalue with
an eigenvector ${\bf z}^*(x)$. Therefore, complex
eigenvalues appear in $(\lambda, -\lambda,\lambda^*, -\lambda^*)$
quadruplets.
Next, if 
\[
{\cal H} {\bf z} = \lambda J {\bf z}, \quad \mbox{Re} \, \lambda \neq 0,
\]
the eigenvector ${\bf z}$ satisfies an identity
\begin{equation}
({\bf z}^*, J {\bf z})=0,
\label{I}
\end{equation}
which follows from the self-adjointness of the operator
${\cal H}$. In \eqref{I}, we used the notation
\[
({\bf z}_1, {\bf z}_2) = \int [ a_1(x) a_2(x) + b_1(x) b_2(x) ] dx,
\]
where $a_i$ and $b_i$ are the (complex) components of the vector ${\bf
z}_i$, i.e.
\[
{\bf z}_i(x)= \left(
\begin{array}{c}
a_i(x) \\
b_i(x) 
\end{array}
\right),
\quad i=1,2.
\]

Assume now that the operator $J^{-1} {\cal H}$ has an eigenvalue
$\lambda$ with $\mbox{Re} \, \lambda >0$, with the eigenvector ${\bf
z}$. It is not difficult to show
that the quadratic form \eqref{star}
calculated on the function
\begin{equation}
{\bf y}= C {\bf z} + C^* {\bf z}^* 
+ {\tilde C} {\tilde {\bf z}}  + {\tilde C}^* {\tilde {\bf z}}^*,
\label{y}
\end{equation}
where $C$ and ${\tilde C}$ are complex coefficients,
  is either sign-indefinite, or
equals zero. Indeed, substituting \eqref{y} into \eqref{star}
and making use of \eqref{I}, we get
\begin{equation}
\frac12 \delta^2 E [ {\bf y}]= 
2 C {\tilde C} \lambda ({\tilde {\bf z}}, J{\bf z}) + c.c. 
+ C^* {\tilde C} (\lambda + \lambda^*) ({\tilde {\bf z}}, J {\bf z}^*)+
c.c.,
\label{qf}
\end{equation}
where ${\tilde {\bf z}}$ is as in \eqref{tilde} 
and $c.c.$ stands for the complex conjugate of the immediately
preceding term. The expression \eqref{qf} is either zero
or changes its sign under 
$C \to -C$. (This conclusion obviously remains valid if
the eigenvalue $\lambda$ is real.)

Thus if the form $\delta^2 E$ is positive definite 
on the subspace $\cal S$ defined by the constraint 
\eqref{constraint}  --- the condition satisfied if the operator $\cal H$
does not have negative or zero eigenvalues other than the translation
mode
--- the product operator
$J^{-1} {\cal H}$ cannot have eigenvalues with positive real 
part.


\begin{thebibliography}{99}

\bibitem{BPS} 
R. Rajaraman. Solitons and Instantons. North-Holland, Amsterdam, 1982;

Y. Yang.  Solitons in Field Theory and Nonlinear Analysis.
Springer-Verlag, New York (2001);

 N. Manton and P. Sutcliffe. 
 Topological Solitons. Cambridge University
Press, Cambridge (2004);

J. Polchinski. String Theory. Cambridge University Press, Cambridge
(2005);
  
 E. J. Weinberg and P. J. Yi, Phys. Rep. {\bf 438}, 65 (2007)
 
 \bibitem{Bazeia}
  
 
D. Bazeia, R. F. Ribeiro and M. M. Santos, Phys. Rev. D {\bf 54}, 1852
(1996);

 
 D. Bazeia, H. Boschi-Filho and F. A. Brito, JHEP {\bf 4}, 28 (1999);
  
 D. Bazeia and F. A. Brito, Phys. Rev. D {\bf 61}, 105019 (2000);
 
 D. Bazeia, J. Menezes and M. M. Santos, Phys. Lett. B {\bf 521}, 418
 (2001);
 
  A. Alonso Izquierdo, 
M.A. Gonz\'alez Le\'on and J. Mateos Guilarte, Phys. Rev. D {\bf 65}, 085012 (2002)

\bibitem{SW}
 J.M. Speight and R.S. Ward,
 Nonlinearity \textbf{7} 475 (1994)

                                                                                                   												   
\bibitem{Ginsburg} L. N. Bulaevskii and V. L. Ginzburg, 
Sov. Phys. JETP {\bf 18},  530 (1964)

\bibitem{Lajzerowicz2}
J. Lajzerowicz and J. J. Niez, J. Phys. (Paris)  {\bf 40}, L165 (1979)



\bibitem{BWZ} I. V. Barashenkov, S. R. Woodford and E. V.  Zemlyanaya, 
Phys. Rev. Lett. {\bf 90}, 054103 (2003)
  

\bibitem{Straka}  
L. Straka and O. Heczko, J. Appl. Phys. {\bf 93}, 8636 (2003);

V.A. Chernenko, V.A. Lvov, S. Besseghini and Y. Murakami,
Scripta Materialia {\bf 55} 307 (2006)
                                                                                
\bibitem{Ivanov} B.A. Ivanov, A.N. Kichizhiev, and Yu.N. Mitsai.
Sov. Phys. JETP {\bf 75} 329 (1992) 
 
\bibitem{Raj_Wei} R. Rajaraman and E.J. Weinberg, 
Phys. Rev. D {\bf 11}, 2950 (1975)

\bibitem{EM} C. Elphick and E. Meron, Phys. Rev. A {\bf 40} 3226 (1989)

\bibitem{Montonen} C. Montonen, Nucl. Phys. B {\bf 112}, 349 (1976) 


\bibitem{Sar_Tru_Bish} S Sarker, S E Trullinger, and A R Bishop,
Phys. Lett. A {\bf 59}, 255 (1976)

\bibitem{BW}
I V Barashenkov and S R Woodford, Phys. Rev. E {\bf 71} 026613 (2005);

I V Barashenkov and S R Woodford, Phys. Rev. E {\bf 75} 026605 (2007)



\bibitem{GL_mag}
Y. Pomeau, Physica D {\bf 51} 546 (1991);


P. Coullet, J. Lega, and Y. Pomeau,
Europhys. Lett. {\bf 15} 221 (1991);
 
 A. Gordon and R. Salditt, Solid State Commun. {\bf
82} 911 (1992)


\bibitem{Lyapunov1}
G. Iz\'us, R. Deza, O. Ram\'irez, H.S. Wio, D.H. Zanette, C. Borzi,
Phys. Rev. A {\bf 52} 129 (1995);

J A Powell, J. Math. Biol. {\bf 35} 729 (1997)



\bibitem{Lyapunov2}
E W Laedke, K H Spatschek and L Stenflo, 
J. Math. Phys. {\bf 24} 2764 (1983);

E W Laedke and K H Spatschek, In: Differential Geometry, Calculus of
Variations and Their Applications. 
Lecture Notes in Pure and Applied Mathematics, 
vol.{\bf 100}. Editors
G M Rassias and T M Rassias. Marcel Dekker, New York 1985, pp. 335-357;

I V Barashenkov, Phys. Rev. Lett. {\bf 77} 1193 (1996)
   
 

\bibitem{Lajzerowicz1} J Lajzerowicz and J J Niez, In: Solitons and Condensed
Matter Physics. Proceedings of the Symposium on Nonlinear 
(Soliton) Structure and Dynamics in Condensed Matter. Oxford, England,
June 27-29, 1978. Editors A R Bishop and T Schneider.
Springer-Verlag, Berlin 1978, pp.195-198

\bibitem{Ito_Tasaki} H. Ito and H. Tasaki, Phys. Lett. A {\bf 113}
179 (1985)



\bibitem{Skryabin} D.V. Skryabin, A. Yulin, D. Michaelis, 
W.J. Firth, G.-L. Oppo, U. Peschel and F. Lederer,
Phys. Rev. E {\bf 64}, 056618 (2001)


 \bibitem{Sub_Tru} K R Subbaswamy and S E Trullinger, 
 Physica D {\bf 2}, 379 (1981)


\bibitem{Haw_Sab_Tru} P Hawrylak, K R Subbaswamy and S E Trullinger,
Phys. Rev. D {\bf 29}, 1154 (1984)

  \bibitem{Tru_DL}  S. E. Trullinger and R. M. DeLeonardis,
  Phys. Rev. B {\bf 22} 5522 (1980)


\bibitem{Ito} H. Ito, Phys. Lett. A {\bf 112} 119 (1985) 



\bibitem{Guilarte_98}
A Alonso Izquierdo, M A Gonz\'alez Le\'on and J. Mateos Guilarte,
J. Phys. A: Math. Gen. {\bf 31} 209 (1998)
 
 
 \bibitem{GL_gen} 
  P. Coullet, J. Lega, B. Houchmanzadeh and J. Lajzerowicz, 
  Phys. Rev. Lett. {\bf 65}, 1352 (1990);
     
 P. Coullet and K. Emilsson, Physica D {\bf 61}, 119 (1992);
 
K. Staliunas, Journ. Mod. Opt. {\bf 42}, 1261 (1995);
  
S. Longhi, Opt. Lett. {\bf 21}, 860 (1996);
 
 L.S. Tsimring and I.S. Aronson, Phys. Rev. Lett. {\bf 79}, 213 (1997);

V.J. S\'anchez-Morcillo, I. Perez-Arjona, F. Silva, 
G.J. de Valcarcel and E. Roldan, 
Opt. Lett. {\bf 25}, 957 (2000);

 G.-L. Oppo, A.J. Scroggie and W.J. Firth, 
 Phys. Rev. E {\bf 63}, 066209 (2001);
 
 H.-K. Park, Phys. Rev. Lett. {\bf 86}, 1130 (2001);

H.-K. Park and M. B\"ar, Europhys. Lett. {\bf 65}, 837 (2004)


 
 \bibitem{BI_gen} 
 B.A. Malomed and A.A. Nepomnyashchy, Europhys. Lett. {\bf 27}, 649
 (1994);
 
 H. Tutu and H. Fujisaka, Phys. Rev. B {\bf 50} 9274 (1994);

 E. Meron, Discrete Dynamics in Nature and Society {\bf 4}, 217 (2000);
 
 A. Yochelis, A. Hagberg, E. Meron, A.L. Lin, and H.L. Swinney,
 SIAM J. Applied Dynamical Systems {\bf 1} 236 (2002);

H. Tutu, Phys. Rev. E {\bf 67} 036112 (2003);
 
 A. Yochelis, C. Elphick, A. Hagberg, E. Meron,
 Physica D {\bf 199} 201 (2004);
 
 D. Gomila, P. Colet, G.-L. Oppo and M. San Miguel, 
 J. Opt. B: Quantum Semiclass. Opt. {\bf 6}, S265 (2004);

\bibitem{Guilarte_88} J. Mateos Guilarte, Ann. Phys. (NY)
{\bf 188}  307 (1988)

 

\bibitem{NLS_fluid}  J.W. Miles, J. Fluid Mech. {\bf 148}, 451 (1984);

 B. Denardo, W. Wright, S. Putterman, and A. Larraza,
Phys. Rev. Lett. {\bf 64}, 1518 (1990);

W. Chen, L. Lu, and Y. Zhu,  Phys. Rev. E {\bf 71}, 036622 (2005) 



\bibitem{E_vs_stability}
T. Kapitula, P. G. Kevrekidis, B. Sandstede, Physica D {\bf 195} 263
(2004); 

S. Cuccagna, D. Pelinovsky and V. Vougalter, 
Communications in Pure and Applied Mathematics, {\bf 58} 1 (2005);

T. Kapitula, P. G. Kevrekidis, B. Sandstede, Physica D {\bf 201} 199
(2005);


D. E. Pelinovsky, Proc. Roy. Soc. Lond. A {\bf 461} 783 (2005);


D. E. Pelinovsky and P. G. Kevrekidis, To appear in:
Z. angew. Math. Phys. (2008)




\end{thebibliography}
\end{document}